# TOPOLOGICALLY MODULATED ELECTROMAGNETIC SIGNALS AND PREDICATE GATES FOR THEIR PROCESSING

*G.A. Kouzaev*

*Alliance of Technology and Science Specialists of Toronto*
*2-2075 Meadowbrook Rd., Burlington*
*L7P 2A5 ON Canada*
*Tel.: 1 (905) 3318317, e-mail: g132uenf@hotmail.com*

**Abstract**. In the paper an idea to create predicate logic components for electromagnetic signals carrying digital information by their amplitudes and spatial field distributions is proposed. The two parameters play a role of predicate and predicate variable. A circuitry for digital processing the signals is suggested and considered. The results are interested for design predicate engineering for artificial intellect computers.

**1. Introduction.** The most efficient results in signal processing have been achieved by using optimal logical system for certain type of signals. For instance, Boolean logic is optimal for pulses of a voltage or a current. Quantum logic requires quantum or quantum-like objects [1-3]. Electromagnetic signals have spatial and amplitude characteristics and the question on the optimal set of methods and algorithms for them has not been solved yet. Several years ago a special class of the signals as a series of pulses with analog or digitally changed amplitudes and discretely modulated spatial structures was proposed [4-6]. The spatial characteristics are discrete vector images of electromagnetic fields changed abruptly because their topological nature (Fig.1).

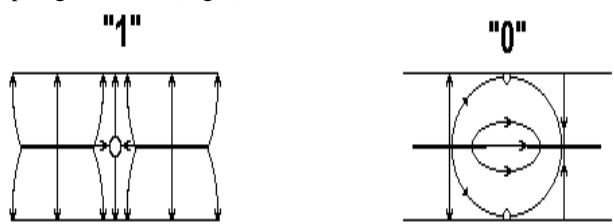

Fig.1. Topological charts of impulses of even and odd modes in a coupled strip transmission line.

Each chart of electromagnetic fields corresponds to a logical variable and transformations of the images are provided due to diffraction effects in passive components of electronic circuits. Theoretically estimated time delay of the transformation is about several parts of a picosecond for micron sized components [4-6]. On Fig. 2 an example of a gate for switching the signals into different layers of IC is shown.

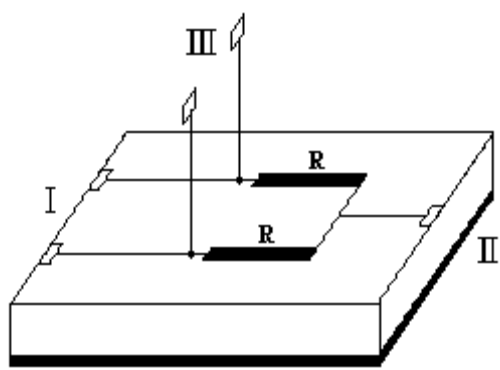

Fig. 2. A passive gate for switching topologically modulated signals into different layers of an IC. I- input coupled strip transmission lines, II – output strip transmission line (even mode impulses, "1"), III- output 2-conductor transmission line (odd mode impulses, "0").

Recent results are in the field of procesing signals by spatal structures and semiconductor gates. Multivalued circuits allowed to process amplitude and spatial information were proposed and modeled for parallel computing. In [6] it was shown a possibility to get the third logical state due to superposition of two topologically modulated fields like in quantum systems. A basic theory of the similarity between quantum logic and logic for spatially modulated signals was considered in [7-9]. In this paper all approaches, including pseudo-quantum gates are united by using predicate logic as a natural sytem for topological signals carrying information by two physical levels.

**2. Application of predicate logic for processing topologically modulated field signals.** As well known, predicate logic is a higher level of logic regarding to the Boolean system [9,10]. The logic is especially pertinent for simulation of natural languages and used in software for

artificial intellect. The simplest predicate logic system *S* consists of a variable *T* and a predicate *A*:
*S=(A, T)*                                                                 (1)

The formula represents a verb construction and means that *variable T has feature of A*. In the case of topologically modulated field signals the amplitudes and spatial structures of them correspond to *A* and *T*, respectively. Thus, the signals are natural carriers of predicate information and that is caused simplicity of fundamental predicate gate design.

**3. Predicate gates for electromagnetic topologically modulated fields.** Predicate logic deals with multiplace signals and allows to process several logical levels at the same time. Due to that the full number of fundamental logical operations depends exponentially on a number of variables and predicates. From one hand, it forces to find a logically non-contradictory set of logical expressions. From another point of view, the large set of logical operation opens a possibility to approximate better the features of natural languages distinguished by their contradictions.
Let's consider a standard set of basic logical operations:
$S = \bar{S}$,
$S = S1 \wedge S2$,
$S = S1 \vee S2$,                                                         (2)
In (2) the first formula means NOT operation, ∧- AND, ∨- OR.
Taking into account manyplaced nature of the signals the full number of logical operations is 45 including atomar formula (1). A logical gate is possible to design for each operation. The design should be based on parrallel processing logical levels contained into different physical characteristic of topologically modulaed signals. Usually, the different nature makes easier the design.

In this paper, the recently developed gates for passive processing topological signals (Fig. 2) were used for design predicate logical circuits. One of the 12 possible gates NOT is shown on Fig. 3 where input 1 and output 6 are coupled strip transmission lines with signals as impulses of odd and even modes .

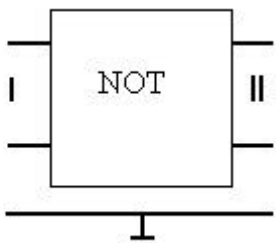

Fig.3. Predicate gate NOT for topologically modulated field pulses. I-input coupled strip transmission lines, II-Output strip transmission lines.

The developed gate inverts only amplitudes of signals thanks to used transistor inverters in the design. As an example on Fig. 4 transient results are shown for inversion input odd mode amplitude. The circuit is able to work with single mode impulses or in multimode regime.

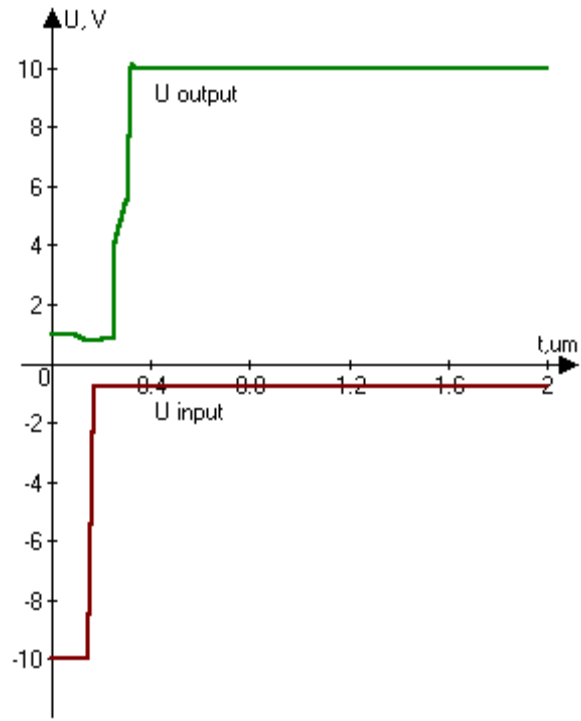

Fig. 4. Input and output diffetrential voltages of an odd mode impulse in the gate NOT.

**4. Conclusion.** It was shown that spatially-modulated electromagnetic signals of certain class allow to realize fundamental predicate gates by an easy manner. The possible circuitry is perspective for design artificial intellect processors.

**References**

1. R.P. Feynman. Quantum mechanical computers. Optics News. 11. (1985). No 2. P. 11.
2. D. Deutch. Quantum theory, the Church-Turing principle and the universal quantum computer. Proc. Royal Soc., A, 400. (1985). P. 97 –117.